\documentclass[aps,prl,notitlepage,twocolumn]{revtex4-1}
\usepackage{amsmath,amssymb,dsfont,graphicx}

\begin{document}

\renewcommand{\k}{{\bf k}}
\newcommand{\bra}[1]    {\left\langle #1\right|}
\newcommand{\ket}[1]    {\left| #1 \right\rangle}
\newcommand{\braket}[2]    {\langle #1 | #2 \rangle}
\newcommand{\braketbig}[2]    {\big\langle #1 \big| #2 \big\rangle}
\newcommand{\braketbigg}[2]    {\bigg\langle #1 \bigg| #2 \bigg\rangle}
\newcommand{\tr}[1]    {{\rm Tr}\left[ #1 \right]}
\newcommand{\av}[1]    {\langle{#1}\rangle}
\newcommand{\avbig}[1]    {\big\langle{#1}\big\rangle}
\newcommand{\avbigg}[1]    {\bigg\langle{#1}\bigg\rangle}
\newcommand{\x}{\mathbf{r}}
\newcommand{\bk}{\mathbf{k}}
\newcommand{\bp}{\mathbf{p}}
\newcommand{\re}{\mathrm{Re}}
\newcommand{\im}{\mathrm{Im}}
\newcommand{\cfi}{F_\mathrm{cl}}

%\title{Correlated atomic pairs scattered from a spinor Bose-Einstein condensate and the violation of Bell inequality}
\title{Bell inequality, Einstein-Podolsky-Rosen steering and quantum metrology with spinor Bose-Einstein condensates}
%\title{Nonclassical atomic pairs from spinor Bose-Einstein condensates for Bell inequality violation, Einstein-Podolsky-Rosen steering and quantum metrology}

\author{T. Wasak and J. Chwede\'nczuk}
\affiliation{Faculty of Physics, University of Warsaw, ul. Pasteura 5, PL--02--093 Warszawa, Poland}

%% \begin{abstract}
%%   We propose an experiment, where the Bell inequality is violated in a many-body system of massive particles.
%%   The source of correlated atoms is a spinor $F=1$ Bose-Einstein condensate residing in an optical lattice.
%%   We characterize the complete experimental procedure--- the local operations, the measurements and the inequality---necessary to run the Bell test. We show how the degree of
%%   violation of the Bell inequality depends on the strengths of the two-body correlations and on the number of scattered pairs.
%%   With the possibility to generalize our analysis to other configurations in a straightforward way, the presented inquiry
%%   can be important in the planning of the forthcoming Bell tests in correlated atomic systems.
%% \end{abstract}
\begin{abstract}
  We propose an experiment, where the Bell inequality is violated in a many-body system of massive particles.  The source of correlated atoms is a spinor $F=1$
  Bose-Einstein condensate residing in an optical lattice.  We characterize the complete experimental procedure--- the local operations, the measurements and the
  inequality---necessary to run the Bell test. We show how the degree of violation of the Bell inequality depends on the strengths of the two-body correlations and on the
  number of scattered pairs.  We show that the system can be used to demonstrate the Einstein-Podolsky-Rosen paradox. Also, the scattered pairs are an excellent
  many-body resource for the quantum-enhanced metrology.  With the possibility to generalize our analysis to other configurations in a straightforward way, the
  presented inquiry can be important in the planning of the forthcoming Bell tests in correlated atomic systems.
\end{abstract}

\maketitle

% somewhere to cite this: dai2016generation

Pairs of entangled particles play an important role in tests of the fundamentals of quantum mechanics.  For example, entangled pairs of photons were used to observe the
Hong-Ou-Mandel effect~\cite{HOM1984measurement}, the ghost imaging~\cite{erkmen2010ghost,shapiro2012physics} or the non-local dispersion
cancellation~\cite{baek2009nonlocal, wasak2010entanglement}.  Recently, the ideas that initially emerged in the domain quantum optics are also applied for massive particles.
The high control over the internal and the external degrees of freedom of quantum gases allows to observe
these phenomena  with matter-waves.  Entangled pairs of atoms were produced in ultracold atomic systems~\cite{boironPhysRevA.87.061603, twin_beam,
  vogels2002generation, perrin2007observation, krachmalnicoff2010spontaneous,truscott}, revealing the sub-Poissonian atom number fluctuations~\cite{jaskula2010sub}, the
violation of the Cauchy-Schwarz inequality~\cite{cauchy_paris, cauchy, cauchy_long}, atomic ghost imaging~\cite{khakimov2016ghost}, or the Hong-Ou-Mandel
effect for helium atoms~\cite{lewis2014proposal,lopes2015atomic}.

Among many quantum phenomena, the violation of the Bell inequality is of particular importance \cite{bell}.
With photons, it was observed with the polarizations~\cite{freedman1972experimental, aspect1981experimental,
PhysRevLett.49.1804,PhysRevLett.49.91, PhysRevLett.61.50, PhysRevLett.81.5039, eibl2003experimental, zhao2003experimental},
and for the phase and momentum entanglement~\cite{PhysRevLett.64.2495}.
In a complex light-matter system, a superposition of a single photon
was transferred into the entanglement of two distant massive objects, which violated a Bell inequality \cite{loophole}.
The Bell correlations were observed with
up to 14 ions in a Paul trap~\cite{lanyon2014experimental}, with spins of either two ions in a single trap~\cite{rowe2001experimental} or two
independently trapped atoms~\cite{hofmann2012heralded}, with
solid state spins~\cite{pfaff2013demonstration} or the Josephson phase qubits~\cite{ansmann2009violation}.
The Bell inequality was also violated by a pair of strongly interacting spin-1/2
hadrons generated in a decay of a shortlived ${}^2$He singlet state ~\cite{PhysRevLett.97.150405}.
The quantification of the quantumess of many-body systems with the correlation functions~\cite{Tura1256}
allowed for the theoretical~\cite{PhysRevA.93.022115} and the experimental~\cite{schmied2016bell} predictions for the presence of the Bell-type correlations.

Although Bell tests were done with massive particles, there is a demand to probe the many-body
nonlocality not only by signalling the presence of the Bell-type correlations, but rather by executing a complete Bell sequence with two remote many-body subsystems~\cite{proposal}.
For this purpose, the atomic systems are particularly useful, as atoms can be prepared in configuration, where they do not interact at large separations,
also offering unique tunability in wide range of parameters.

In this work we propose the setup where the Bell inequality violation and the Einstein-Podolsky-Rosen (EPR) steering~\cite{epr} can be demonstrated in a many-body system
of massive partciles~\cite{wasak2016role}. We also prove that the system can be potentially useful for quantum enhanced metrology tasks.
We consider a Bose-Einstein condensate (BEC) in the $F=1,\ m_F=0$ hyper-fine state immersed in the optical lattice
\cite{twin_paris}.  This potential modifies a single particle dispersion relation and triggers the scattering of atomic pairs with opposite angular momentum projections,
$m_F=\pm1$, into two separated regions.  Since there is no communication between these regions, as atoms weakly interact through the contact potential, the resulting
entangled state is well-suited for testing the postulates of local realism \cite{bell}.  

Recently, a similar setup, but without referring to spin states, was used to show the atomic Hong-Ou-Mandel effect~\cite{lopes2015atomic} and observe a two-particle
interference of a pair of atoms with entangled momenta~\cite{dussarrat2017two}. The setup considered here is highly controlable: the presence of an optical lattice triggers 
the dynamical instability, while the number of atoms can be tailored by tuning the duration of the optical lattice potenial~\cite{twin_paris}.  The
possibility to tune the occupations of the modes, to generate beams of atoms with controllable velocities, and to prepare many-body entangled states makes the proposed
setup a complementary approach to experiments with trapped atoms in optical lattice, as in~\cite{dai2016generation}, where a violation of CHSH-type
inequality was observed between two atoms in a double well of a deep optical superlattice.  The presented inquiry contributes to the developing field of testing the quantum mechanics with
matter waves \cite{PhysRevA.91.052114,keller2014bose,kofler2012epr,peise2016satisfying,gneiting2008bell}.

We provide an analysis of the Bell problem for this system: we propose the local operations, the measurement and the Bell inequality.  We derive an experimentally
important relation, which links the degree of violation of the Bell inequality with the number of scattered pairs and the magnitude of the two-body correlations. We show
that the Bell inequality will be violated if the number of scattered atoms is of the order of unity. It imposes severe requirements on the sensitivity of the single-atom
detection, but this level of precision has already been demonstrated \cite{dall2013ideal,khakimov2016ghost}.  When the number of scattered atoms is large, the system will
not violate the Bell inequality, though still will be of great value for other non-classical tasks, such as the demonstration of the EPR paradox or quantum-enhanced
metrology \cite{wasak2014bogoliubov}.  

To analyze the scattering, we use the field  operator
$\hat\Psi_\alpha(\x)$, which destroys an atom with the spin projection $\alpha=0,\pm1$ at $\x$,
and satisies the commutation relation
${[} \hat\Psi_\alpha^{\phantom{\dagger}}(\x), \hat\Psi_{\alpha'}^\dagger(\x')  {]}  = \delta_{\alpha\alpha'} \delta(\x-\x')$.
The many body Hamiltonian with the two-particle contact interactions
reads (we use the Einstein summation convention)
\begin{subequations}
\begin{eqnarray}
  &&\hat H = \int\!\! d\x\,\bigg[
      \frac{\hbar^2}{2m} \nabla\hat\Psi^\dagger_\alpha(\x) \cdot \nabla\hat\Psi_\alpha(\x)\\
      && + V(\x) \hat\Psi_\alpha^\dagger(\x)\hat\Psi_\alpha(\x)+ \frac{c_0}{2} \hat\Psi_\alpha^\dagger(\x)
      \hat\Psi_\beta^\dagger(\x) \hat\Psi_\beta(\x) \hat\Psi_\alpha(\x)\\
      &&+ \frac{c_1}{2} \hat\Psi_\alpha^\dagger(\x) \hat\Psi_\beta^\dagger(\x)  \mathbf{F}_{\alpha,\alpha'} \cdot \mathbf{F}_{\beta,\beta'}
             \hat\Psi_{\beta'}(\x) \hat\Psi_{\alpha'}(\x)\bigg].
\end{eqnarray}
\end{subequations}
Here, $m$ is the atomic mass, while $c_0=\frac{4\pi \hbar^2 }{3m}(a_0 + 2 a_2)$ and $c_1=\frac{4\pi \hbar^2 }{3m}(a_2 - a_0)$ are the strengths of the two-body interactions;
the scattering lengths $a_0$ and $a_2$ are calculated for the total angular momentum $J=0$ or $J=2$ channels, respectively. By dot we denote contractions of eulicdean tensors.
Finally, $V(\x)$ combines the harmonic trap and the optical lattice, and $\mathbf{F}$ is a vector of spin-1 matrices.

If the number of scattered particles is small, the Bogoliubov approximation applies. It treats the fields of scattered atoms
$\hat\delta_\alpha$ as
small perturbations, i.e., up to linear in the equations of motion, on top of the source BEC $\phi_\alpha$ \cite{dalfovo1999theory}.
Each component $\phi_\alpha$ satisfies the coupled mean-field equations:
\begin{eqnarray}\label{gpe}
  i \hbar \partial_t \phi_\alpha &=& \bigg[ -\frac{\hbar^2}{2m}\nabla^2 + V \bigg]\phi_{\alpha}\nonumber
  + c_0 \bigg(\sum_\beta |\phi_{\beta}|^2 \bigg)\phi_{\alpha}\\
  &+& c_1 \bigg( \phi_{\alpha'}^* \mathbf{F}_{\alpha',\beta'} \phi_{\beta'}  \bigg) \cdot \mathbf{F}_{\alpha,\beta} \phi_{\beta},
\end{eqnarray}
while $\hat \delta_\alpha$ operators are governed by the time-dependent Bogoliubov equations
\begin{eqnarray}\label{bog}
  &&i\hbar\partial_t \hat\delta_\alpha \nonumber=\bigg( - \frac{\hbar^2}{2m}\nabla^2 + V + c_0 \sum_\beta |\phi_\beta|^2 \bigg)\hat\delta_{\alpha} \\
  &&+\bigg\{ c_0 \phi^*_{\beta}\phi_\alpha + c_1 \bigg[ (\mathbf{F}_{\alpha,\beta'} \cdot \mathbf{F}_{\alpha',\beta} + \mathbf{F}_{\alpha,\beta}\mathbf{F}_{\alpha',\beta'} )
  \phi_{\alpha'}^* \phi_{\beta'} \bigg]  \bigg\} \hat\delta_{\beta}\nonumber\\
    && + \bigg\{ c_1 \phi_{\beta}\phi_\alpha + c_1  \mathbf{F}_{\alpha,\alpha'} \cdot \mathbf{F}_{\beta,\beta'} \phi_{\alpha'} \phi_{\beta'}   \bigg\} \hat\delta_{\beta}^\dagger.
\end{eqnarray}
Initially, all atoms reside in the $m_F=0$ state.  The addition of the optical lattice drives the BEC into the dynamical instability and triggers the scattering of atoms
in pairs out of the condensate. This process is energetically forbidden in the absence of the lattice, thus by tuning the duration of lattice, one may control the number
of scattered atoms.

\begin{figure}[t!]
  \includegraphics[clip, scale=0.24]{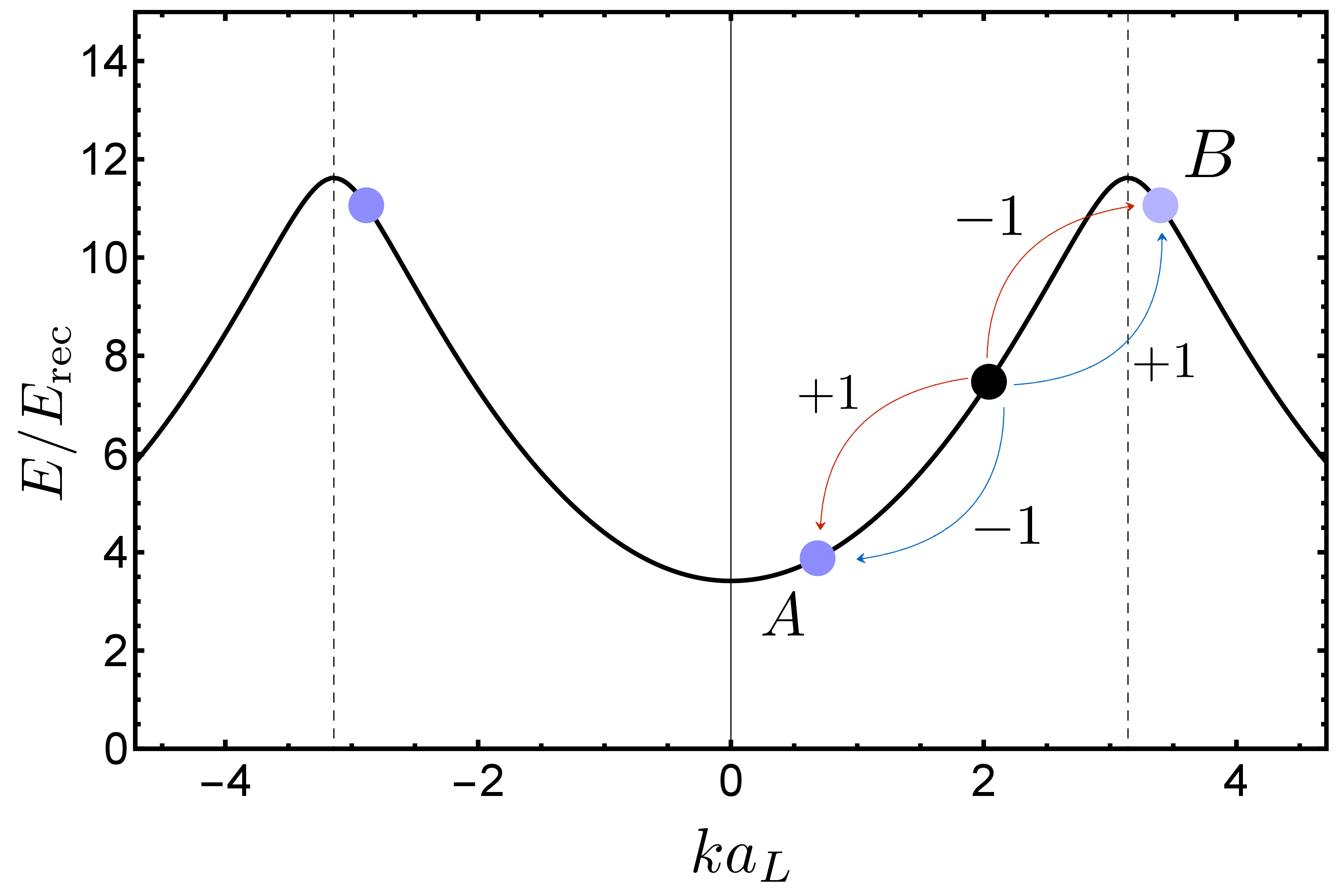}
  \caption{(Color online) The energy of atoms in the lattice in units of the recoil energy $E_{rec}=\frac{\hbar^2\pi^2}{2ma_L^2}$ (solid black line).
    The BEC with the momentum $ka_L=2.04$ is the black dot.
    The solution of the Schr\"odinger equation predicts the scattering of an atomic pair out of the BEC, represented by violet dots at $ka_L=0.686$ and $ka_L=-2.885$ (this
    one is moved to the neighbouring Brillouin zone for clarity). A process where the scattering is accompianied by the spin flip, results in the emission of the atom pairs: either $m_F=1$ to $A$ and
    $m_F=-1$ to $B$ (red arrows) or coherently vice-versa (blue arrows). The parameters used to prepare this plot were taken from \cite{twin_paris} and are discussed in detail in the results section.
    The horizontal dashed lines separate Brillouin zones.}
  \label{scheme}
\end{figure}

The momenta of the outgoing pair are found with the one-body Schr\"odinger equation (cf. Eq.~(\ref{gpe}) with $c_0 = c_1 = 0$):
the wave-function $\phi_{0}(x)$ is decomposed into the Fourier series $\phi_{0}(x)=\sum_nC_ne^{i\frac{2\pi}{a_L}x\cdot n}$, where $a_L$ is the lattice
period and the numerical solution of the coupled equations for the $C_n$'s gives the dispersion relation $E(k)$.
If the BEC is at rest or moves slowly with respect to the lattice, the conservation laws
for the momenta of the incoming and the outgoing pair, $E(k^{\rm in}_1)+E(k^{\rm in}_2)=E(k^{\rm out}_1)+E(k^{\rm out}_2)$ and $k^{\rm in}_1+k^{\rm in}_2=k^{\rm out}_1+k^{\rm out}_2$, can be satisfied only if the outgoing
pair remains inside the BEC. Above the critical value of the BEC momentum, pairs scatter out of the the condensate.
The collision of two BEC atoms can either keep $m_F$'s of the outgoing pair unchanged or, more interestingly, result in the spin-flipping scattering of a $m_F=\pm1$ pair (denoted from
now on as $\uparrow/\downarrow$),
which after the free expansion is spatially separated into two regions $A$ and $B$, see Fig.~\ref{scheme}. Therefore, the scattering process will yield a state in the form
\begin{equation}\label{state}
  \ket{\psi(t)}=\mathcal{C}_0(t)\ket0+\mathcal{C}_2(t)\Big(\ket{\uparrow_A,\downarrow_B}+\ket{\downarrow_A,\uparrow_B}\Big)+\ldots
\end{equation}
The dots denote the omitted, and intially irrelevant contribution from a higher number of pairs, while the amplitudes $\mathcal{C}_i(t)$
are obtained by solving the coupled equations (\ref{gpe}) and (\ref{bog});
note that in this simplified notation we do not specify the structire of the momentum modes in this multi-mode problem.
Clearly, the state (\ref{state}) has the Bell-type two-atom component.
In order to verify if the contribution from the vacuum or from the states populated by a
higher number of pairs would not spoil the desired effect, we describe in detail the complete experimental Bell sequence.

To run the Bell test and sample the coherent superposition of Eq.~(\ref{state}), the optical lattice is turned off and 
the local operations independently in $A$ and in $B$ mix the states $\uparrow$ and $\downarrow$.
Thus we introduce the many-body equivaltents of the spin-$1/2$ Pauli matrices which operate within each region, i.e., 
\begin{subequations}
  \begin{eqnarray}
    \hat J_x^{\alpha} &=& \frac12 \int_\alpha\!\!\frac{d\k}{2\pi}\bigg( \hat\delta_{\uparrow}^\dagger(\bk)\hat\delta_{\downarrow}(\bk) + \hat\delta_{\downarrow}^\dagger(\bk)\hat\delta_{\uparrow}(\bk)  \bigg),\label{jx} \\
    \hat J_y^{\alpha} &=& \frac{1}{2i} \int_\alpha\!\!\frac{d\k}{2\pi}\bigg( \hat\delta_{\uparrow}^\dagger(\bk)\hat\delta_{\downarrow}(\bk) - \hat\delta_{\downarrow}^\dagger(\bk)\hat\delta_{\uparrow}(\bk)  \bigg), \label{jy}\\
    \hat J_z^{\alpha} &=& \frac12 \int_\alpha\!\!\frac{d\k}{2\pi}\bigg( \hat\delta_{\uparrow}^\dagger(\bk)\hat\delta_{\uparrow}(\bk) - \hat\delta_{\downarrow}^\dagger(\bk)\hat\delta_{\downarrow}(\bk)  \bigg), \\
    \hat N_{\alpha} &=& \frac12 \int_\alpha\!\!\frac{d\k}{2\pi}\bigg( \hat\delta_{\uparrow}^\dagger(\bk)\hat\delta_{\uparrow}(\bk) + \hat\delta_{\downarrow}^\dagger(\bk)\hat\delta_{\downarrow}(\bk)  \bigg),
  \end{eqnarray}
\end{subequations}
where $\alpha = A,B$ denotes the integration over one of the regions.
They satisfy the commutation relations $[\hat J_i^{\alpha},\hat J_j^{\alpha'}]= i \epsilon_{i,j,k} \hat J_k^{\alpha}\delta_{\alpha,\alpha'}$,
$[\hat N_\alpha , \hat J_i^{\alpha} ]=[(\hat J^{\alpha})^2 , \hat J_i^{\alpha}] = 0$. For instance, (\ref{jx}) is an analog of the $\hat \sigma_x$ Pauli operator---it transfers a particle
residing in region $\alpha$ from $\uparrow$ to the $\downarrow$ and vice-versa. Physically, transformations generated by these operators can be performed by RF-coupling the two hyper-fine levels~\cite{lucke2011twin}.

We consider the following Bell sequence. The state (\ref{state}) is locally and independently in $A$ and in $B$ transformed by one of the above operators, here we choose
the $y$-axis rotation for illustration. As a result, we obtain
\begin{equation}\label{trans}
  \ket{\theta,\phi}=e^{-i\theta\hat J_y^{(A)}}e^{-i\phi \hat J_y^{(B)}}\ket{\psi(t)},
\end{equation}
where $\theta$ and $\phi$ are the two rotation angles. After the transformation is made, the atom number difference between the two hyper-fine levels is measured in $A$ and $B$ and the results
are correlated and normalized to give
\begin{equation}
  E(\theta,\phi) =\frac{\bra{\theta,\phi} \hat J_z^{(A)} \hat J_z^{(B)} \ket{\theta,\phi} }{ \bra{\theta,\phi}\hat N_A \hat N_B\ket{\theta,\phi}}.
\end{equation}
According to \cite{reid}, the Bell inequality for this quantity is
\begin{equation}\label{be}
    \mathcal{B}\equiv\Big|E(\theta,\phi) + E(\theta',\phi') + E(\theta',\phi) - E(\theta,\phi')\Big|\leqslant2,
\end{equation}
i.e., the procedure requires four setting of the angles: ($\theta$, $\theta'$) in $A$ and ($\phi$, $\phi'$) in $B$.

\begin{figure}[t!]
  \includegraphics[clip, scale=0.37]{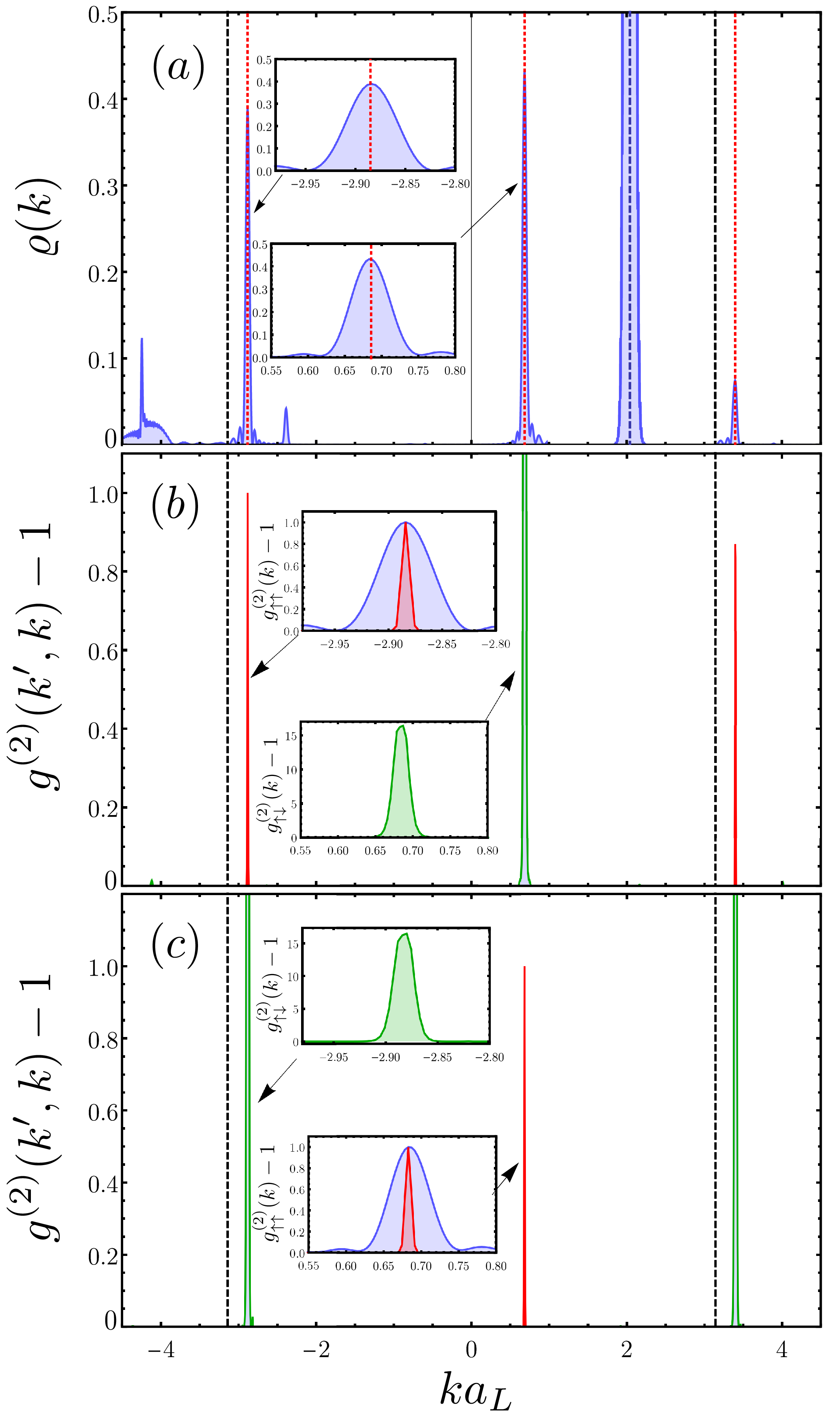}
  \caption{(Color online) The density and correlations of the scattered atoms at $t=0.7$~ms. 
    (a) The density $\varrho(k)=\av{\hat \delta^\dagger_\uparrow(k)\hat \delta_\uparrow(k)}$ (in the harmonic oscillator units in $z$-direction) 
    in the $\uparrow$-component obtained from the numerical solution of the Bogoliubov equation, and the mother BEC (at position $ka_L=2.04$). The vertical dashed red
    lines are the resonant momenta predicted by the solutions of the one-body Schr\"odinger equation. The inset shows the details of the density of each cloud. 
    % At that time we have 0.29 particles in region A (left peak) and 0.35 in region B (right peak).
    (b) and (c) The second order correlation function $g^{(2)}(k)$ with $k'$ fixed in $A$, $k'a_L=-2.885$ (b)
    and in $B$, $k'a_L=0.686$ (c). The red represents the local-, while the green the back-to-back second order coherence.}
  \label{fig.dens}
\end{figure}

The quantum state (\ref{state})  derived from the Bogoliubov equations (\ref{bog}) is characterized by its first- and  second-order correlation functions.
Since the initial state of the scattered pairs is a vacuum, the correlation $E(\theta,\phi)$
can be expressed analytically in the following form \cite{supp}:
\begin{equation}
  E(\theta,\phi) = \frac{\av{\hat J_z^{A}\hat J_z^{B}}}{\av{\hat N_A\hat N_B}} \cos(\theta+\phi).
\end{equation}
Here and below the average is calculated in the state $\ket{\psi(t)}$, i.e., prior to the local operations.
This expression substituted into Eq.~(\ref{be}) and optimized over the angles yields
\begin{equation}\label{opt}
  \max_{\mathrm{angles}}\mathcal{B}=2\sqrt2 \bigg|\frac{\av{\hat J_z^A\hat J_z^B}}{\av{\hat N_A\hat N_B}}\bigg|.
\end{equation}
Therefore, the system will violate the Bell inequality if
\begin{equation}\label{will}
  \bigg|\frac{\av{\hat J_z^A\hat J_z^B}}{\av{\hat N_A\hat N_B}}\bigg|>\frac1{\sqrt 2}.
\end{equation}
For the Bogoliubov scattering into symmetric and disjoint regions, this last equation can be re-written as \cite{supp}
\begin{equation}\label{cond2}
  \bigg|\frac{\av{\hat J_z^A\hat J_z^B}}{\av{N_AN_B}}\bigg|=\frac{g^{(2)}_{\uparrow\downarrow} - 1}{ g^{(2)}_{\uparrow\downarrow} + 1},
\end{equation}
where the normalized back-to-back second order correlation function reads
\begin{equation}\label{g2}
  g^{(2)}_{\uparrow\downarrow}=\frac{\iint\limits_{A\,B} d\k d\k' \av{\hat \delta^\dagger_{\uparrow}(\k)\hat \delta^\dagger_{\downarrow}(\k')\hat \delta_{\downarrow}(\k')\hat \delta_{\uparrow}(\k)}}
  {\iint\limits_{A\,B} d\k d\k'\av{\hat \delta^\dagger_{\uparrow}(\k)\hat \delta^{\phantom\dagger}_{\uparrow}(\k)}\av{\hat \delta^\dagger_{\downarrow}(\k')\hat \delta^{\phantom\dagger}_{\downarrow}(\k')}}.
\end{equation}
We combine Eqs~(\ref{will}) and (\ref{cond2}) and conculde that the Bell inequality will be violated when
\begin{equation}\label{cond}
  g^{(2)}_{\uparrow\downarrow}>2\sqrt2 + 3.
\end{equation}
This condition is of particular experimental interest as it determines the minimal value of the cross-region correlation for which the Bell inequality may potentially be violated.

We now solve Eqs~(\ref{gpe}) and (\ref{bog}) using the experimental parameters of \cite{twin_paris} and show that the condition (\ref{cond}) is met in this many-body system: 
a BEC of $N=10^4$ ${}^4$He$^*$ atoms
\footnote{It should be noted that ${}^4$He$^*$ can suffer from inelastic losses when not spin-polarized. The magnetic moment is large and the mass is low, so this species requires a good control over the stray fields. (Chris Westbrook, private communication)}.
We take a cylindrical trap with the frequencies $(\nu_x,\nu_y,\nu_z)=(1500,1500,6.5)$Hz. The optical lattice is a 1D potential,
$U_0\sin^2(k_{\rm rec}x)$, with the wave-vector $k_{\rm rec}=5.9\mu{\rm m}^{-1}$, recoil energy $E_{\rm rec}=44$kHz$\cdot h$ and $a_L=\pi/k_{\rm rec}=0.53\mu m$.  The
depth of the lattice is $U_0=0.725E_{\rm rec}\equiv{\hbar^2 k_0^2}/{2m}$.  For scattering lengths we take $\bar a=(a_0+2a_2)/3=6.77$nm and $\Delta a=(a_2-a_0)/3 =0.73$nm. 
The BEC moves with the momentum $k_0a_L=2.04$ with respect to the lattice and the one-body Schr\"odinger equation predicts that atomic pairs should scatter into the
momenta $k_1a_L=0.686$ and $k_2a_L=-2.885$, as shown schematically in Fig.~\ref{scheme}. With these parameters, we numerically solve the Bogoliubov equations
(\ref{bog}). The Fig.~\ref{fig.dens}(a) shows the density of the scattered atoms at $t=0.7$ms, together with the mother BEC. Clearly the position of the peaks of scattered
clouds coincide with the predictions.

Next, we calculate the second order correlation function $g^{(2)}_{\uparrow\downarrow}(k,k')$, given by Eq.~(\ref{g2}) before the integration over the regions.
We fix one variable in $A$ (at  $k'a_L=-2.885$)  or in $B$
(at $k'a_L=0.686$), see Fig.~\ref{fig.dens}(b) and (c), and scan $k$ throught the system. The presence of strong cross-correlation between $A$ and $B$ (green lines)
confirms that indeed the scattering is a two-body process.

\begin{figure}[t!]
  \includegraphics[clip, scale=0.24]{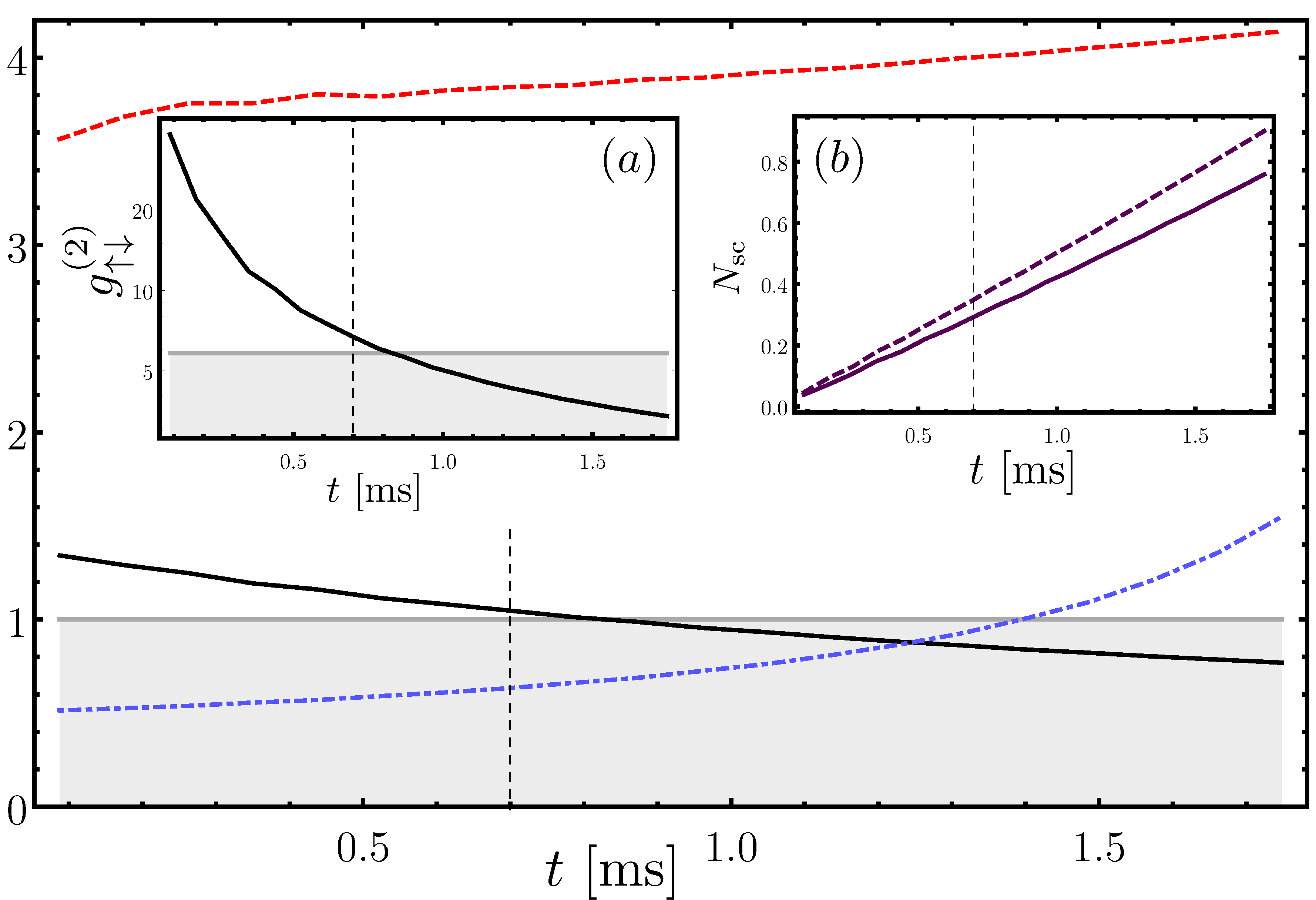}
  \caption{(Color online) Main figure: signatures of non-classicality as a function of time. All quantites are normalized, so that their values above unity (outside the
    gray shaded area) signify the quantumness of the atomics pairs.  The solid black line is the Bell coefficient $\mathcal B/2$ defined in Eq.~(\ref{opt}). The
    dot-dashed blue line is the inverse of the Hillery-Zubairy coefficient, i.e., $\frac12E^{-1}_{\rm HZ}$, while the dashed red line is the QFI normalized to the shot-noise
    level.  The left inset (a) shows the value of the $g^{(2)}_{\uparrow\downarrow}$, while the right inset (b) displays the average number of scattered atoms in $A$
    (solid line) and in $B$ (dashed line).  The vertical dashed lines indicate the instant of time at which the density and the correlations shown in Fig.~\ref{fig.dens}
    were calculated.}
  \label{fig.bell}
\end{figure}

Finally, we calculate the Bell coefficient $\mathcal{B}$ with Eq.~(\ref{opt}) and plot the result as a function of time in Fig.~\ref{fig.bell}.  The system does violate
the Bell inequality up to $t\sim0.85$ ms. The insets show that the strength of the cross-correlation function drops from the inital value $\gg1$ to the threshold
$2\sqrt2+3$ only at the last instants of the dynamics. The number of scattered pairs is low (on average around 1/2 particle per $A$ and $B$), which keeps the system deep
in the quantum regime. This is consistent with the property of the Bogoliubov system, which have a positive-defined (and thus semi-classical) Wigner function when the
number of scattered atoms is large \cite{single}.  However, at later times the system is still much quantum, as witnessed by other entanglement criteria. To demonstrate
this, we calculate the Hillery-Zubairy coefficient
\begin{align}
  E_{\rm HZ}=1+\frac{\av{\hat J_+^A\hat J_-^A\hat J_+^B\hat J_-^B}-\left|\av{\hat J_+^A\hat J_-^A}\right|^2}{2\,{\rm min}\left[\av{\hat J_z^A\hat J_+^B\hat J_-^B},\av{\hat J_+^A\hat J_-^A\hat J_z^B}\right]},
\end{align}
where $\hat J_{\pm}^\alpha=\hat J_x^\alpha\pm i\hat J_y^\alpha$.  The values $E_{\rm HZ}<\frac12$ indicate the strongly non-classical EPR steering entanglement
\cite{epr,steering,steering2,steering3}. The Fig.~\ref{fig.bell} shows that this type of correlation emerges in the system at later times, when the Bell coefficient drops
below the quantum-to-classical threshold. We also calculate the quantum Fisher information (QFI) for the interferometric transormation, which imprints the phase $\theta$
between the two spin components in $A$ and $-\theta$ in $B$. The QFI is given by
\begin{align}
  F_q=4\left(\av{\hat J_z^2}-\av{\hat J_z}^2\right), 
\end{align}
where $\hat J_z=\hat J_z^A-\hat J_z^B$. When $F_q>\av{\hat N_B+\hat N_B}$, i.e., exceeds the shot-noise level, the system is particle-entangled and useful for
quantum-enhanced metrology \cite{braunstein1994statistical,giovannetti2004quantum,pezze2009entanglement}. The Fig.~\ref{fig.bell} shows that the QFI monotonically grows
with time, indicating that the scattered atomic pairs are an excellent resource for the sub shot-noise interferometry \cite{wasak_twin}.

%% \begin{figure}[t!]
%%   \includegraphics[clip, scale=0.24]{FigBellViolation_new.eps}
%%   \caption{(Color online) Violation of the Bell inequality defined in Eq.~(\ref{be}) as a function of time. The classical $\mathcal{B}=2$ value is crossed only at the final times of the evolution.
%%     The right inset shows the value of the $g^{(2)}_{\uparrow\downarrow}$, while the left inset displays the average number of scattered atoms in $A$ (solid line) and in $B$ (dashed line).
%%   The vertical dashed lines indicate the instant of time at which the density and the correlations shown in Fig.~\ref{fig.dens} were calculated.}
%%   \label{fig.bell}
%% \end{figure}

In conclusion, we have presented a complete sequence for testing a Bell inequality in a many-body system of massive particles. The $F=1$ BEC in an optical lattice is a
good source of correlated $\uparrow/\downarrow$ pairs scattered into two separeted region. Local operations mix the projections of the spin and the subsequent measurement
of the number of atoms in each component are the basic building blocks for the Bell inequality. We have derived the analytical expression for the Bell coefficient
$\mathcal B$ and related its value to the strength of the back-to-back two-body correlation function, which is of the particular experimental relevance. The analytical
results are backed by the numerical solution of the Bogoliubov equations using the experimental parameters of \cite{twin_paris}. The presented results will find
application in designing the experiemnts testing the Bell inequalites.
An open issue is if the Bell inequality can be violated by a nonclassical two-region state  with multimode many-body structure,
which is highly-populated. 
However, as our study shows, at such stage, the system is still useful for other non-classical tasks, such as the demonstration of the EPR paradox or the quantum-enhanced metrology.
Finally, we remark that the theory presented here applies to other setups where correlated pairs
scatter from a coherent source. For a free-space collision of two $m_F=0$ BEC's, our results apply directly. If one of the two colliding condensates is in another spin
state, our findings can be used without any modifications for as long as the mean-field repulsion is small compared to the characteristic kinetic enegy of the system,
i.e., the ``skier effect'' is negligible \cite{skier}.

We thank Andrew Truscott, Chris Westbrook and Denis Boiron for valuable discussions and for carefully reading the manuscript. 
This work was supported by the Polish Ministry of Science and Higher Education programme ``Iuventus Plus'' for years
2015-2017, project number IP2014 050073, and the National Science Center Grant No. 2014/14/M/ST2/00015.

\section{Appendix}

\subsection{Bogoliubov equations}
When the condensate resides in the $\alpha=0$ state, the GPE (equation (2) from the main text) simplifies to  $i \hbar \partial_t \phi_0 = \bigg[ -\frac{\hbar^2}{2m}\nabla^2 + V + c_0  |\phi_{0}|^2\bigg]\phi_{0}$
and the Bogoliubov equations read
\begin{subequations}
\begin{eqnarray}
  &&i\hbar \partial_t \hat\delta_{0} = \hat H_{00} \hat\delta_{0} + c_0 \phi_0^2 \hat\delta_{0}^\dagger\\
  &&i\hbar \partial_t \hat\delta_{\pm} = \hat H_{0} \hat\delta_{\pm}\pm c_1 \phi_0^2 \hat\delta_{\pm}^\dagger,\label{bogs}
\end{eqnarray}
\end{subequations}
where the symmetric/anti-symmetric combinations of the $\pm1$ fields (from now on denoted as $\uparrow/\downarrow$ for the consitiency of the notiation) are 
$\hat\delta_{\pm}=\frac{1}{\sqrt{2}}\bigg( \hat\delta_{\uparrow}\pm \hat\delta_{\downarrow} \bigg)$.
The Hamiltonians are given by the formulas
\begin{subequations}
\begin{eqnarray}
  &&\hat H_{00} = -\frac{\hbar^2}{2m} \nabla^2 + V + 2c_0|\phi_0|^2\\
  &&\hat H_{0} = -\frac{\hbar^2}{2m} \nabla^2 + V + (c_0+c_1)|\phi_0|^2.
\end{eqnarray}
\end{subequations}
We focus only on the dynamics of the $\uparrow/\downarrow$ fields. Since the Equations (\ref{bogs}) are linear, their solution is
\begin{equation}\label{sol}
  \hat\delta_{\pm}(\x,t) = \int\!\!\!d\x \bigg[ C_{\pm}(\x,\x'|t)\hat \delta_{\pm}(\x') + S_{\pm}(\x,\x'|t)\hat \delta_{\pm}^\dagger(\x') \bigg],
\end{equation}
where for brevity $\hat \delta_{\pm}(\x,0)\equiv\hat \delta_{\pm}(\x)$, while 
 $C_{\pm}(\x,\x'|0) = \delta(\x-\x')$ and $S_{\pm}(\x,\x'|0) = 0$. 
We now discuss the properties of the system in detail.

\subsection{Correlation functions}

The first-order correlation function and the anomalous density in the momentum coordinates are
\begin{equation}\label{coh}
  G_{s,s'}(\bk,\bk') = \av{ \hat\delta_{s}^\dagger(\bk) \hat\delta_{s'}(\bk') },\  M_{s,s'}(\bk,\bk') = \av{ \hat\delta_{s}(\bk) \hat\delta_{s'}(\bk') },
\end{equation}
where $s,s'=\pm1$. This can be expressed in terms of the coherence functions of the fields $\pm$ to give
\begin{subequations}
\begin{eqnarray}
  &&G_{s,s'}(\bk,\bk') = \frac12 \bigg[ G_{+}(\bk,\bk') + (ss') G_{-}(\bk,\bk')\bigg]\\
  &&M_{s,s'}(\bk,\bk') = \frac12 \bigg[ M_{+}(\bk,\bk') + (ss') M_{-}(\bk,\bk')\bigg],
\end{eqnarray}
\end{subequations}
which holds since the Eq.~(\ref{bogs}) for $\hat\delta_+$ and $\hat\delta_-$ do not couple, while
$G_{\pm}(\bk,\bk') = \av{ \hat\delta_{\pm}^\dagger(\bk) \hat\delta_{\pm}(\bk') }$ and $M_{\pm}(\bk,\bk') = \av{ \hat\delta_{\pm}(\bk) \hat\delta_{\pm}(\bk') }$.
The second order correlation function  is
\begin{equation}
  G^{(2)}_{s_1,s_2,s_3,s_4}(\bk_1,\bk_2,\bk_3,\bk_4) = \av{ \hat\delta_{s_1}^\dagger(\bk_1) \hat\delta_{s_2}^\dagger(\bk_2) \hat\delta_{s_3}(\bk_3) \hat\delta_{s_4}(\bk_4)}.
\end{equation}
For the Bogloiubov system, the Wick's theorem applies, which allows to express the $G^{(2)}$ in terms of the lower-order correlations from Eq.~(\ref{coh}) as follows
\begin{widetext}
  \begin{equation}\label{2nd}
    G^{(2)}_{s_1,s_2,s_3,s_4}(\bk_1,\bk_2,\bk_3,\bk_4)=M_{s_1,s_2}^*(\bk_1,\bk_2)M_{s_3,s_4}(\bk_3,\bk_4)
    + G_{s_1,s_3}(\bk_1,\bk_3)G_{s_2,s_4}(\bk_2,\bk_4)
    + G_{s_1,s_4}(\bk_1,\bk_4)G_{s_2,s_3}(\bk_2,\bk_3).
  \end{equation}
\end{widetext}

\subsection{Bell inequality}
For the Bell sequence, we take the local operations generated by $\hat J_y^\alpha$, thus 
\begin{equation}
  \hat U_A(\theta)\hat U_B(\phi) = e^{i \theta\hat  J_y^A }e^{i \phi\hat J_y^B}.
\end{equation}
After the transformations, the atom number difference between the $\uparrow$ and the $\downarrow$ states
is measured, and the results are correlated to give
\begin{equation}
  E(\theta,\phi) = \frac{ \av{\hat J_z^A(\theta)\hat J_z^B(\phi) } }{\av{\hat N_A\hat N_B}},
\end{equation}
The Bell inequality is
\begin{equation}
  |E(\theta,\phi) + E(\theta',\phi') + E(\theta',\phi) - E(\theta,\phi')|^2\equiv |\mathcal{B}|\leqslant2.
\end{equation}
We now proceed to calculate the correlation function $E(\theta,\phi)$ for the spinor system. 
The local rotations transform the operators as $\hat J_z^\alpha(\theta) = \hat J_z^\alpha \cos\theta + \hat J_x^\alpha \sin\theta$, giving
\begin{widetext}
\begin{equation}
  E(\theta,\phi) = \frac{ \av{\hat J_z^A\hat J_z^B} \cos\theta\cos\phi+\av{\hat J_z^A\hat J_x^B} \cos\theta\sin\phi+\av{\hat J_x^A\hat J_z^B} \sin\theta\cos\phi+\av{\hat J_x^A\hat J_x^B} \sin\theta\sin\phi  }
  {\av{\hat N_A\hat N_B}}.
\end{equation}
\end{widetext}
These averages express by the $G^{(2)}$ from Eq.~(\ref{2nd}) as
\begin{widetext}
\begin{subequations}
  \begin{eqnarray}
    \av{\hat J_z^A \hat J_z^B} &=& \iint\limits_{A\,B}\!\!\frac{d{\bf k}d{\bf p}}{16\pi^2}
    \bigg[G_{+}^{*}(\bk,\bp) G_{-}(\bk,\bp) +  G_{+}(\bk,\bp)G_{-}^{*}(\bk,\bp)+ M^*_{+}(\bk,\bp)M_{-}(\bk,\bp) + M_{+}(\bk,\bp)M_{-}^*(\bk,\bp)\bigg],\\
    \av{\hat J_x^A \hat J_x^B}&=&\iint\limits_{A\,B}\!\!\frac{d{\bf k}d{\bf p}}{16\pi^2}
    \bigg[|G_{+}(\bk,\bp)|^2 +  |G_{-}(\bk,\bp)|^2 + |M_{+}(\bk,\bp)|^2 +  |M_{-}(\bk,\bp)|^2+\nonumber\\
      &+&(G_{-}(\bk,\bk)-G_{+}(\bk,\bk))(G_{-}(\bp,\bp)-G_{+}(\bp,\bp))\bigg],\\
    \av{\hat N_A \hat N_B} &=&\iint\limits_{A\,B}\!\!\frac{d{\bf k}d{\bf p}}{16\pi^2}
    \bigg[|G_{+}(\bk,\bp)|^2 +  |G_{-}(\bk,\bp)|^2 + |M_{+}(\bk,\bp)|^2 +  |M_{-}(\bk,\bp)|^2+\nonumber\\
      &+&(G_{-}(\bk,\bk)+G_{+}(\bk,\bk))(G_{-}(\bp,\bp)+G_{+}(\bp,\bp))
      \bigg],
  \end{eqnarray}
\end{subequations}
\end{widetext}
while $\av{\hat J_x^A \hat J_z^B}=\av{\hat J_z^A \hat J_x^B}=0$.
Since $\bk \in A$ and $\bp \in B$ (thus they lie in the two disjoint regions),
the first order coherence among the regions vanishes, $G_{+/-}(\bk,\bp) = 0$. Also, using the
symmetry $G_{+}=G_{-}$ and $M_{+}=-M_{-}$ (resulting directly from Equations (\ref{bogs})), we obtain
\begin{eqnarray}
  \av{\hat J_z^A\hat J_z^B}&=& - \iint\limits_{A\,B}\!\!\frac{d{\bf k}d{\bf p}}{8\pi^2}|M_{-}(\bk,\bp)|^2,\\
  \av{\hat N_A\hat N_B} &=&  \iint\limits_{A\,B}\!\!\frac{d{\bf k}d{\bf p}}{8\pi^2} \bigg[ |M_{-}(\bk,\bp)|^2+2 G_{-}(\bk,\bk) G_{-}(\bp,\bp)\bigg]\nonumber
\end{eqnarray}
and $\av{\hat J_x^A\hat J_x^B}=-\av{\hat J_z^A\hat J_z^B}$, while
\begin{equation}
  E(\theta,\phi) = \frac{\av{\hat J_z^A\hat J_z^B}}{\av{\hat N_A\hat N_B}}\cos(\theta+\phi).
\end{equation}
The Bell coefficient  maximized over the angles is
\begin{equation}
  \max_{\mathrm{angles}}{\mathcal B} = 2\sqrt2 \bigg|\frac{\av{\hat J_z^A\hat J_z^B}}{\av{\hat N_A\hat N_B}}\bigg|.
\end{equation}
In the final step by further exploit the symmetry of the system 
$M_{-} = - M_{\uparrow\downarrow} = - M_{\downarrow\uparrow}$ and  $G_{\pm}=G_{\uparrow\uparrow} = G_{\downarrow\downarrow}$,
which gives
\begin{equation}
  G^{(2)}_{\uparrow\downarrow\uparrow\downarrow}(\bk,\bp,\bk,\bp)=|M_{\uparrow\downarrow}(\bk,\bp)|^2 + G_{\uparrow\uparrow}(\bk,\bk)  G_{\downarrow\downarrow}(\bp,\bp) 
\end{equation}
The integration over regions $A$ and $B$ yields
\begin{equation}
  G^{(2)}_{\uparrow\downarrow}=\iint\limits_{A\,B}\!\!\frac{d{\bf k}d{\bf p}}{4\pi^2}\bigg[ |M_{\uparrow\downarrow}(\bk,\bp)|^2+ G_{\uparrow\uparrow}(\bk,\bk)  G_{\downarrow\downarrow}(\bp,\bp)\bigg].
\end{equation}
Finally, we introduce the normalized second order correlation function
\begin{equation}
  g^{(2)}_{\uparrow\downarrow} = \frac{ G^{(2)}_{\uparrow\downarrow}}{\iint\limits_{A\,B}\!\!\frac{d{\bf k}d{\bf p}}{4\pi^2}\,G_{\uparrow\uparrow}(\bk,\bk)  G_{\downarrow\downarrow}(\bp,\bp)  }.
\end{equation}
which reads
\begin{equation}
  g^{(2)}_{\uparrow\downarrow}= 1 + \frac{\iint\limits_{A\,B}\!\!\frac{d{\bf k}d{\bf p}}{4\pi^2}|M_{\uparrow\downarrow}(\bk,\bp)|^2  }{
    \iint\limits_{A\,B}\!\!\frac{d{\bf k}d{\bf p}}{4\pi^2}\,G_{\uparrow\uparrow}(\bk,\bk)  G_{\downarrow\downarrow}(\bp,\bp)},
\end{equation}
giving
\begin{equation}
  \max_{\mathrm{angles}}{\mathcal B}=2\frac{g^{(2)}_{\uparrow\downarrow}-1}{ g^{(2)}_{\uparrow\downarrow}+ 1},
\end{equation}
which is the central result of our work.

\end{document}